# Perspective of spintronics applications based on the Extraordinary Hall Effect.

A. Gerber and O. Riss

Raymond and Beverly Sackler School of Physics and Astronomy,

Tel Aviv University,

Ramat Aviv, 69978 Tel Aviv, Israel

Extraordinary Hall effect (EHE) is a spin-dependent phenomenon that generates voltage proportional to magnetization across a current carrying magnetic film. Magnitude of the effect can be artificially increased by stimulating properly selected spin-orbit scattering events. Already achieved sensitivity of the EHE-based sample devices exceeds 1000 $\Omega$/T, which surpasses the sensitivity of semiconducting Hall sensors. Linear field response, thermal stability, high frequency operation, sub-micron dimensions and, above all, simplicity, robustness and low cost manufacture are good reasons to consider a wide scale technological application of the phenomenon for magnetic sensors and memory devices.

Anisotropic magnetoresistance [1,2], planar Hall effect [3,4], spin-dependent tunneling [5,6] and the extraordinary or anomalous Hall effect (EHE) [7,8] are spin-dependent electronic transport phenomena known for many years. However, it is the discovery of the giant magnetoresistance (GMR) [9,10] that gave birth to the term spintronics and triggered a world-wide outburst of the spin-related research. Extraordinary Hall Effect (EHE) in magnetic materials was discovered more than a century ago [7], extensively studied both theoretically and experimentally [8], and left out of the mainstream research for the last thirty years. The possibility to use the effect for technical applications, such as magnetic sensors and nonvolatile magnetic random access memories (MRAM), has been mentioned more than three decades ago [11], but no significant progress was reported until recently. A probable reason for this is that although EHE in bulk magnetic materials can be significantly higher than the ordinary Hall effect in normal metals, its magnitude remained far beyond the sensitivity of semiconductors and magnetic sensors based on the anisotropic magnetoresistance [12]. The renewed interest in EHE has only recently arisen when some recipes to enhance the effect were found [13-15].

The Hall effect in magnetic materials is commonly described [8,16] by the phenomenological equation

$$\rho_H = R_0 B + R_{EHE} \mu_0 M = R_0[H + \mu_0 M(1-D)] + R_{EHE} \mu_0 M \tag{1}$$

where $\rho_H$ is the Hall resistivity, $B$, $H$ and $M$ are components of the magnetic induction, applied field and magnetization normal to the film plane, and $D$ is the demagnetization factor. $R_0$ is the ordinary Hall coefficient related to the Lorentz force acting on moving charge carriers. $R_{EHE}$, the extraordinary Hall coefficient, is associated with a break of the right-left symmetry at spin-orbit scattering in magnetic materials.. Demagnetization factor D is equal to 1 when field is applied perpendicular to a homogeneous magnetic film. In this case Eq.1 is simplified to

$$\rho_H = R_0 H + R_{EHE} \mu_0 M \tag{2}$$

Voltage measured between Hall contacts located perpendicular to the direction of an electric current is given by:

$$V_H = \frac{I}{t}\rho_H = \frac{I}{t}(R_0 H + R_{EHE}\mu_0 M) \quad (3)$$

where $I$ is current and $t$ thickness of the film.

Fig.1 presents a typical field dependence of the Hall resistance ($R_H \equiv V_H/I$) measured in a 4 nm thick Ni film at room temperature with field applied normal to the film plane. Magnetization normal to the film increases with field till saturation at about ±0.2 T. The EHE contribution is constant at higher fields, and further minor variation of the Hall resistance is contributed by the ordinary Hall effect. In cases of our interest the EHE contribution exceeds significantly the ordinary Hall effect term in the low field range, and Hall voltage $V_H$ can be approximated as:

$$V_H = R_H I = \mu_0 R_{EHE} M I / t \quad (4)$$

Sensitivity of the Hall resistance to external field can be defined as:

$$S = \frac{dR_H}{dH} = \frac{\mu_0 R_{EHE} \chi}{t} \quad (5)$$

where $\chi$ is magnetic susceptibility ($\chi = dM/dH$). Sensitivity is a function of three parameters: thickness $t$, the EHE coefficient $R_{EHE}$ and susceptibility $\chi$. All three can be used to enhance the field sensitivity in thin ferromagnetic films.

It is generally accepted that EHE in ferromagnetic metals originates from the spin-orbit scattering that breaks a spatial symmetry in the trajectory of scattered electrons. Since scattering is responsible, both for EHE and longitudinal resistivity, a link between two parameters is usually claimed. Two types of scattering events are distinguished in the EHE literature. One is referred to as skew scattering and is characterized by a constant spontaneous angle at which the scattered carriers are deflected from their original trajectories [17]. The predicted [8, 18] correlation between the EHE coefficient and resistivity ρ is $R_{EHE} = A\rho + B\rho^2$. The second term is frequently neglected and a linear ratio between $R_{EHE}$ and ρ is mentioned. The other scattering mechanism, so-called side jump [19], is quantum in nature and results in a constant lateral displacement of the charge trajectory at the point of scattering. For the side jump process $R_{EHE} \propto \rho^2$, and this mechanism is expected to dominate in highly resistive samples at elevated temperatures or with high degree of doping. In general, experimental data does not fit well the predictions of the models. It was then suggested [20] that in systems with several different scattering sources, the total resistivity is not a good parameter to characterize the effect. Instead, both the EHE and resistivity should be decomposed to contributions generated by different scattering sources and the correlation for each source should be followed independently. Surfaces, impurities, and thermal scattering contributions, analyzed separately, were shown [20] to follow the skew scattering model regardless of the total material's resistivity. An additional mechanism connected with Berry phase and believed to take place in the absence of any scattering is also discussed [21]. Comprehensive understanding of the phenomenon has not yet been achieved; however, a link between spin-orbital scattering, resistivity and EHE coefficient provides a guideline towards improvement of the EHE efficiency. Here we shall review three cases: effects of the interfaces and surface scattering, scattering by impurities distributed in ferromagnetic host and scattering by ferromagnetic clusters embedded in conducting non-magnetic matrices.

**Enhancement of sensitivity by surface scattering.**

Following the original Fuchs size-effect model [22,23], external surfaces impose a boundary condition on the electron-distribution function, which enhances the intrinsic, thickness-independent bulk resistivity $\rho_b$ to a thickness-dependent resistivity $\rho$. Resistivity of thin films increases when the mean free path becomes comparable with thickness and diverges in the zero-thickness limit. Increase of resistivity by surface scattering was shown [14] to lead to the respective increase in the EHE coefficient. Longitudinal resistivity $\rho$ and Hall resistivity $\rho_H$ of a series of Ni films are shown in Fig. 2 as a function of thickness. Both parameters diverge in the thin film limit. One can define the surface scattering contribution to resistivity of a given film as a difference between resistivity of the sample and resistivity of thick (bulk) samples: $\rho_{ss} = \rho - \rho_b$. In a similar way one can define the EHE component contributed by surface scattering as: $\rho_{Hss} = \rho_H - \rho_{Hb}$ The inset of Fig. 2 reveals a linear correlation between $\rho_{Hss}$ and $\rho_{ss}$.

Similar results were found in Co monolayers, Co-Pd bilayers [24], Fe-Pt films [25] and Fe/Ge multilayers [26]. Fig.3 presents the room temperature field sensitivity of polycrystalline Ni films as a function of their thickness. Increase of sensitivity from 20 m$\Omega$/T in 100 nm thick film to 30 $\Omega$/T in 3 nm thick film is due both to the reduction of the thickness *t* and to the increase of the EHE resistivity in the thin film limit. Enhanced EHE resistivity and field sensitivity is a general property of ultrathin ferromagnetic materials observed also in CoFe/Pt multilayers [27], thin amorphous Fe-Ge [28] and ultrathin Fe-Pt films [25].

**Insulating impurities in ferromagnetic host.**

Spin-orbit scattering by adatoms and grains of insulating materials dispersed in ferromagnetic metals is a source of a strong extraordinary Hall effect. Fig. 4 presents a linear correlation between the insulator scattering contributions to EHE resistivity and linear resistivity, found in granular Ni-SiO$_2$ mixtures with a relatively low concentration of SiO$_2$ [14]. At higher concentrations of SiO$_2$ (about 50% depending on deposition conditions [29,30] the system approaches the metal – insulator transition with resistivity diverging above 1$\Omega$cm and the saturated Hall resistivity reaching 200 $\mu\Omega$cm [13]. For 1 $\mu$m thick films the field sensitivity is about 10 $\Omega$/T. The effect was first reported by Pakhomov et al [13] and was called the Giant Hall effect. Similar values were found in other granular ferromagnet-insulator systems, like Co- SiO$_2$ [31], NiFe-SiO$_2$ [32] and Fe-SiO$_2$ [33].

**Magnetic clusters in non-magnetic metallic host.**

Granular films with ferromagnetic grains embedded in a normal metal matrix form a different class of systems with important interface spin-orbit scattering. The EHE generated in this type of materials is proportional to magnetization of ferromagnetic grains, and as such can serve an extremely sensitive tool to monitor the properties of nano-scale magnetic objects [34]. In granular ferromagnets EHE coefficient can be by orders of magnitude larger than in bulk materials due to an effective interface scattering. Fig. 5 presents a typical Hall signal measured in a planar array of uniform 3 nm size Co clusters embedded in Pt matrix as a function of the applied field. The samples were prepared by the low-energy clusters beam deposition (LECBD) technique [35]. Blocking temperature of Co clusters is below 100 K, therefore the magnetization and Hall resistance have no hysteresis at room temperature. The extraordinary Hall coefficient and, respectively, the sensitivity to an applied field increase linearly with the density of magnetic clusters (see inset of Fig. 5). As a result, the sample with just 0.7 nm effective thickness of Co (which corresponds to an inter-particle distance of about 4.4 nm) shows the field sensitivity of 7.4 Ω/T. Planar arrays of Co nanoparticles embedded in W matrix were reported [36] to show sensitivity of 32 Ω/T. These values are by orders of magnitude higher than in bulk cobalt.

**Effect of an out-of-plane anisotropy.**

Saturation field is another important parameter affecting the susceptibility $\chi$, sensitivity and range of applications of the EHE-based magnetic sensors. Saturation is achieved at the demagnetization field $H_s = 4\pi M_s$, where $M_s$ is the saturated magnetization, when field is applied perpendicular to an isotropic ferromagnetic film. The saturation field can be reduced dramatically in the presence of an out-of-plane anisotropy. Fig. 6 illustrates the effect of anisotropy on the field dependent Hall resistance of 5 nm thick Ni films. Surface induced out-of-plane anisotropy dominates at low temperatures, and Hall resistance shows a square hysteresis with sharp reversal of magnetization at the switching field. Geometrical in-plane anisotropy prevails above the reorientation phase transition temperature. Hall resistance at room temperature is hysteresis-free and saturates at the demagnetization field [37]. Magnetization reversal in the vicinity of the reorientation phase transition is very sharp due to the development of the multi-domain structure with the out-of-plane anisotropy. [38,39]. As a result, the Hall resistance close to the reorientation transition (see the 207 K curve in Fig. 6) reverses sharply yet has no hysteresis. The field sensitivity of this sample increases from 7 Ω/T at room temperature to 275 Ω/T at 175 K in the reversible range and up to 530 Ω/T at 60 K in the irreversible range.

Multilayered ferromagnet-normal metal films with an out-of-plane anisotropy can be used to reduce the saturation field and enhance the EHE sensitivity at ambient temperature. Fig. 7 presents the field dependent Hall resistance measured in a thin Co-Pd multilayer sample at room temperature [40]. Reversal of magnetization is accomplished below 20 Oe, and the EHE sensitivity exceeds 300 Ω/T. Further improvement of sensitivity is possible by tuning the composition and thickness. Sensitivity as high as 1200 Ω/T was recently reported in thin CoFe/Pt multilayers [41].

Sensitivity of the order of 1000 Ω/T is in the range of the best ever achieved in magnetic sensors. It is justified, therefore, to spend more effort and test other characteristics of the possible EHE devices.

**Hysteresis-free operation.**

Hysteresis-free operation is implemented in systems with hysteresis-free magnetization, like thin ferromagnetic films with field applied normal to the easy anisotropy axis or granular superparamagnetic systems above their blocking temperature. Examples of the hysteresis-free operation are illustrated in Figs.1, 5 and 7.

**Dynamic range.**

Dynamic range depends on material and geometry. As mentioned earlier, the highest saturation field of a continuous isotropic film with field applied normal to its plane is the demagnetization field given by $H_{sat} = 4\pi M_s$. The operation range of materials depends on the value of their saturated magnetization. For isotropic Ni and Fe films these are about ±0.4 T and ±1.6 T respectively (see Fig.8). Operation range of paramagnetic and superparamagnetic systems can be extended to much higher fields, although at the expense of linearity of the response.

**Linear field response.**

Linear response over significant field range is an important advantage of the EHE-based sensors. Examples of the linear field dependence of the EHE resistance in thin Ni, Co and Fe films are shown in Fig. 8.

**Thermal stability and drift.**

It is quite usual to observe a certain drift of data when resistance of thin magnetic films is measured repeatedly during multiple field cycling. Variation of resistance due to the drift can be comparable and even exceed the entire magnetoresistance values. Example of such drift is shown in Fig. 9a. The origin of the effect is not well understood, but it can be related to unstable temperature, aging or field training of the material. Hall resistance measured simultaneously with the longitudinal resistance is shown in Fig. 9b. No drift is found in the Hall resistance, which is entirely reproducible within the measurement accuracy.

In general, an enhanced EHE signal is obtained in materials with an intensive spin-orbit scattering and high resistivity: amorphous metals, thin films, metals doped by insulating impurities. All these materials are bad metals, and a relative contribution of phonon scattering both in resistivity and EHE is small. As such, the temperature dependence of EHE is weak and tunable by an extent of disorder. Temperature independent sensitivity over a wide temperature range was found in Fe-Pt alloy thin films [42,43] and in amorphous Fe-Ge films [28].

**Dimensions.**

The active element of the EHE sensor is the intersection of the Hall bar between the current carrying line and voltage probes. Scaling of length and width of the junction does not change its longitudinal and Hall resistances. We tested the effect of the Hall bar dimensions on performance of the EHE sensors in a series of lithographically patterned Ni films. Fig.10 presents the EHE response of two Ni samples with Hall bar intersections of 10μm×10μm and 0.1μm×0.1μm. Both samples demonstrate the same field dependence and sensitivity. We can, therefore assume that the size of the EHE devices can be miniaturized to the lithography limits [44].

**High frequency operation.**

Magnetization of soft ferromagnetic thin films [45] and multilayer films with the out-of-plane anisotropy [46] can be operated at GHz frequencies. Since EHE signal is an electrical replica of magnetization, these frequencies are available for the EHE devices. Notably, the frequency range of semiconducting Hall sensors and devices is limited to the MHz range.

**EHE semiconductor spintronics.**

One of the major efforts in the current spintronics research is in attempt to incorporate the spin-dependent degree of freedom and contemporary semiconductor technology. Generation of conducting, ferromagnetic at room temperature semiconductors is mandatory to construct semiconductor spintronics devices operating by the GMR principle. The goal is not yet accomplished due to a limited solubility of magnetic impurities in semiconductor hosts. Materials with relatively high concentration of magnetic impurities are superparamegnetic rather than ferromagnetic. On the other hand, the very existence of ferromagnetic semiconductors is not needed if EHE is adopted as a basis for the semiconductor spintronics. Segregation of phases and creation of superparamagnetic materials with ferromagnetic clusters embedded in a conducting semiconductor host is similar to the all-metallic superparamagnetic systems with a non-magnetic metallic matrix. Spin dependent scattering of charge carriers by magnetic grains generates the EHE signal proportional to magnetization of clusters when the latter are embedded within a conducting matrix, metallic or semiconducting. Fig.11 presents a large EHE signal developed in Mn doped Ge with sensitivity of about 240 Ω/T at 77 K. Similar to the all-metallic superparamagnetic systems, the Hall resistance is hysteretic below and or hysteresis-free above the blocking temperature of ferromagnetic clusters. Hysteresis is observed in the EHE signal at room temperature when magnetic anisotropy of clusters is large and their blocking temperature exceeds the room temperature.

**Memory.**

Magnetic anisotropy is not directly related to the EHE effect; however both features can be correlated in e.g. thin films. As mentioned above, $R_{EHE}$ coefficient is enhanced by diffusive surface scattering in thin films. Simultaneously, the out-of-plane anisotropy due to surface or interface effects can overcome the geometrical in-plane anisotropy resulting in the out-of-plane magnetic easy axis in thin films. As a result one can produce materials with large EHE signal and hysteresis in the field response, suitable for magnetic memory devices. The principle is shown in Fig.12 for Ni films of different thicknesses that possess the out-of-plane magnetic anisotropy at 4.2 K. Magnitude of the effect increases with the reduction in films' thickness. Square hysteresis loop with about 4 Ω separation between the upward and downward magnetized states is obtained in 5 nm thick film. Efficient memory units can be constructed using the same approach with materials showing the out-of-plane anisotropy at room temperature. An example is Co-Pd multilayer sample with a strong EHE signal and a square hysteresis loop at room temperature, shown in Fig. 13. Separation between the up and down magnetized states in this sample exceeds 0.3 Ω, and the switching field is about 230 Oe. Magnitude of the effect and the switching field can be tunable by a proper selection of composition, thickness of each component and a total thickness. Arrays of such units can operate as nonvolatile magnetic memory devices.

**Simplicity.**

One of the major advantages of the EHE-based approach is the structural simplicity and robustness of the devices. Currently developed spintronics devices are generally based on the giant magnetoresistance phenomenon (GMR) in which resistance of a heterogeneous magnetic system is sensitive to a distribution of magnetic moments. GMR devices typically consist of at least three active ferromagnet / non-magnetic spacer / ferromagnet layers but are, in fact, rather complicated multilayered structures. Efficiency of the metallic GMR devices decreases when a number of multilayers drops below the order of a hundred [47]. EHE devices can have a single-layer active element only, which makes their production simple and cheap. It should also be noted that Hall resistance does not depend on the length of the element and is therefore independent of the total resistance of the device. Resistance can be adjusted to the rest of the electronic circuit without compromising the performance of the EHE element.

**Challenges**.

Optimistic picture sketched above is based on limited preliminary data. Application-oriented EHE research is, for the moment, much too premature to point out the fundamental limitations and difficulties of the approach.

**Summary.**

Anisotropic magnetoresistance (AMR) and associated with it planar Hall effect in homogeneous magnetic films and spin-dependent magnetoresistance of heterogeneous ferromagnet/non-magnetic structures (GMR or TMR - type) are two well-known spin-dependent phenomena used to transform magnetic data into electric signals. Extraordinary Hall effect (EHE) is another spin-dependent phenomenon that generates voltage linearly proportional to magnetization across a current carrying film. Sensitivity of EHE can be artificially increased by stimulating properly selected spin-orbit scattering events. Already achieved sensitivity of the EHE sample devices exceeds 1000 $\Omega$/T, which surpasses the sensitivity of semiconducting Hall sensors. Linear field response, thermal stability, high frequency operation, sub-micron dimensions and, above all, simplicity, robustness and low cost manufacture are good reasons to consider a wide scale technological application of the phenomenon for magnetic sensors and memory devices. The technique can be incorporated with modern semiconducting technology by using the natural segregation of magnetic dopants instead of fighting against it.

**Acknowledgements.**

This work was supported in part by the Israel Science Foundation grant No. 633/06. The effort was also sponsored by the Air Force Office of Scientific Research, Air Force Material Command, USAF, under grant number FA8655-07-1-3001. The U.S. Government is authorized to reproduce and distribute reprints for Government purpose notwithstanding any copyright notation thereon. The views and conclusions contained herein are those of the author and should not be interpreted as necessarily representing the official policies or endorsements, either expressed or implied, of the Air Force Office of Scientific Research or the U.S. Government.

# References.

# Figure captions

Fig.1. Hall effect resistance as a function of field applied perpendicular to 4 nm thick Ni film at room temperature.

Fig.2. Resistivity and EHE resistivity of thin Ni films as a function of thickness . Inset: correlation between the surface scattering contributions to EHE resistivity and longitudinal resistivity.

Fig.3. EHE sensitivity of thin Ni films as a function of thickness. Room temperature.

Fig.4. Correlation between the increase of EHE resistivity and longitudinal resistivity induced by $SiO_2$ impurities in Ni at 77K (open circles) and room temperature (solid circles). Straight line is guide for the eyes.

Fig.5. Hall resistance measured in a planar array of Co clusters embedded in a thin Pt matrix as a function of applied magnetic field. Inset: EHE resistivity as a function of Co clusters density. Diameter of Co clusters is 3 nm..

Fig.6. Hall resistance of 5 nm thick Ni film as a function of applied field at 4.2 K, 207 K and 300 K. Magnetic anisotropy easy axis is out-of-plane at low temperature and in-plane at room temperature. Magnetic field is applied perpendicular to the film's plane.

Fig. 7. Room temperature Hall resistance of a thin Co-Pd multilayer with a tuned out-of-plane anisotropy. Saturation field is below 20 Oe. Sensitivity is 300 $\Omega$/T.

Fig.8. Hall resistance of 6 nm thick Co and 5 nm thick Fe films as a function of field normal to the films plane. T = 290 K. The response is linear below the saturation fields.

Fig.9. Longitudinal resistance (a) and Hall resistance (b) of a thin Ni film measured several times by repeating field cycling at room temperature. Note the drift in resistance, its magnitude comparable with the total magnetoresistance changes. No drift is observed in the Hall resistance data.

Fig. 10. Hall response of two Ni films with different cross-sections of the Hall bars (10μm ×10μm) and (0.1μm × 0.1μm). Sub-micronization of the device does not affect its sensitivity.

Fig.11. Hall resistance measured in the Mn implanted Ge sample. The material is superparamagnetic. EHE sensitivity is about 250 Ω/T at 77 K.

Fig. 12. Memory-type hysteretic response of Ni films of different thickness at 4.2 K. Separation between the up and down magnetized states in 5 nm thick sample is about 4 Ω.

Fig. 13. Illustration of the room temperature magnetic memory unit based on the EHE. Shown is the EHE resistance of the Co-Pd multilayer sample with the out-of-plane anisotropy at room temperature. Separation between the up and down magnetized states exceeds 0.3 Ω. The switching field of this sample is about 230 Oe.

**Figures.**

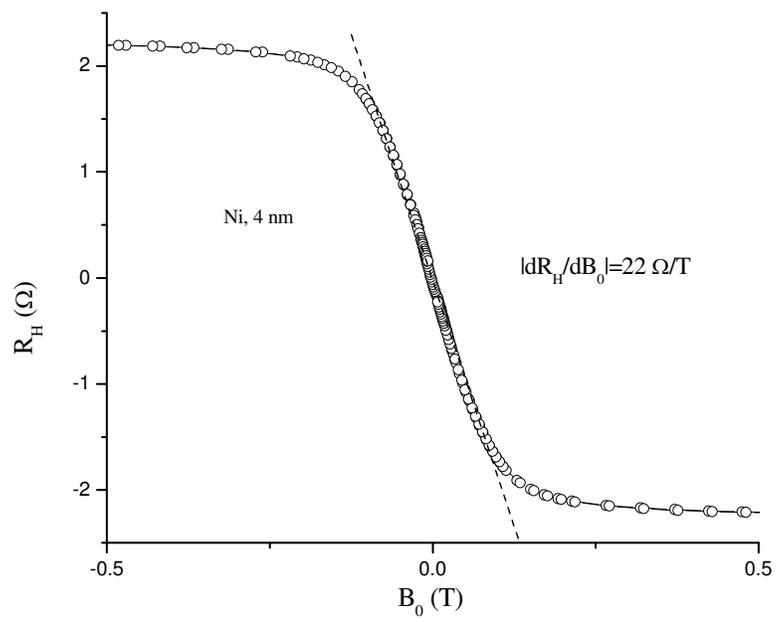

Fig.1

Fig.2

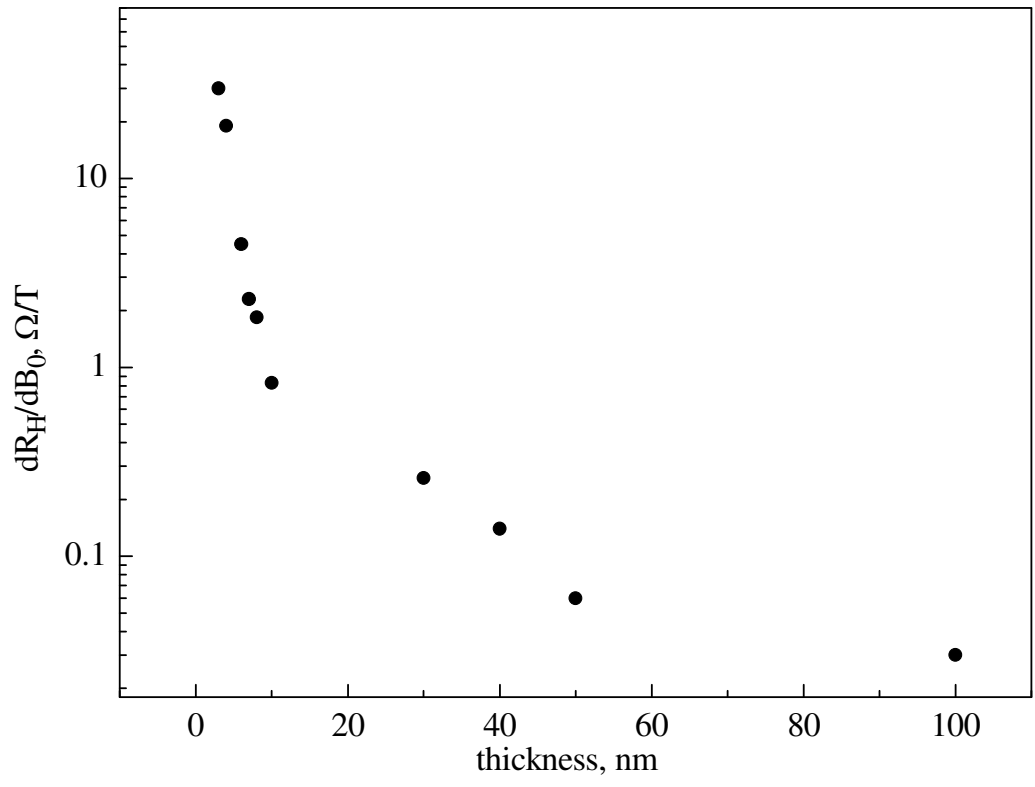

Fig.3

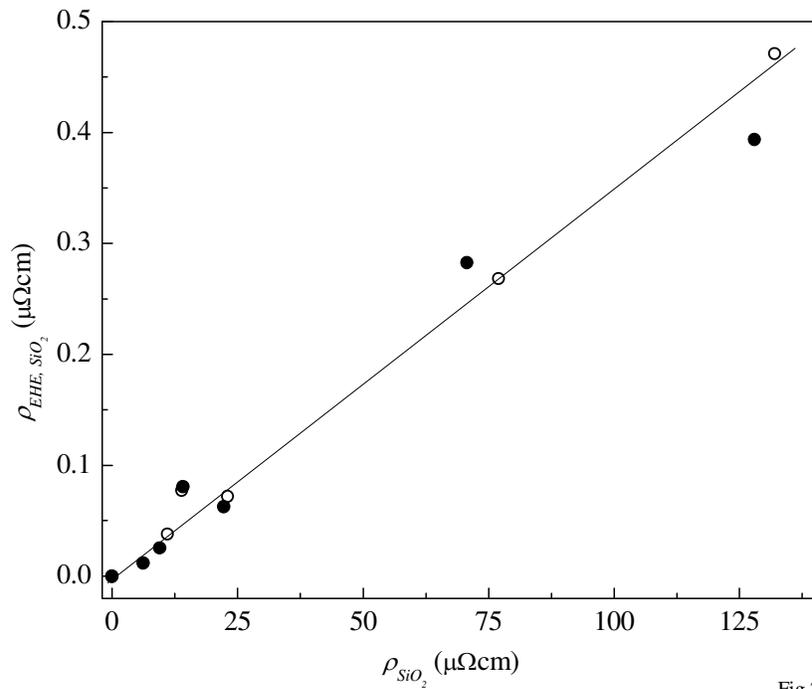

Fig. 4

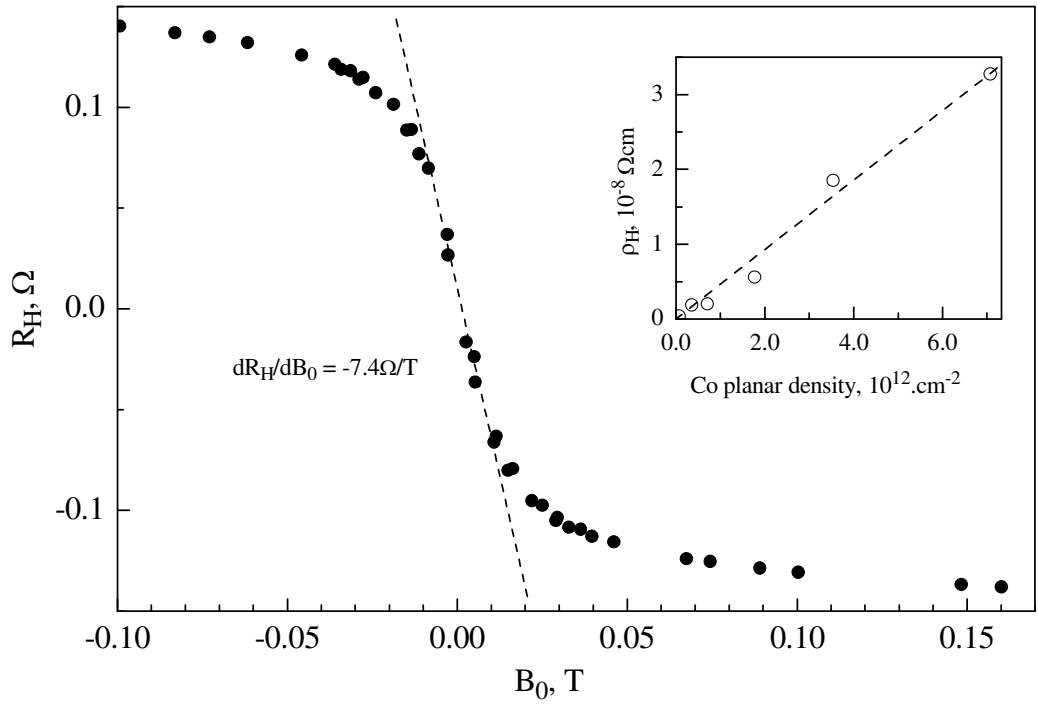

Fig.5

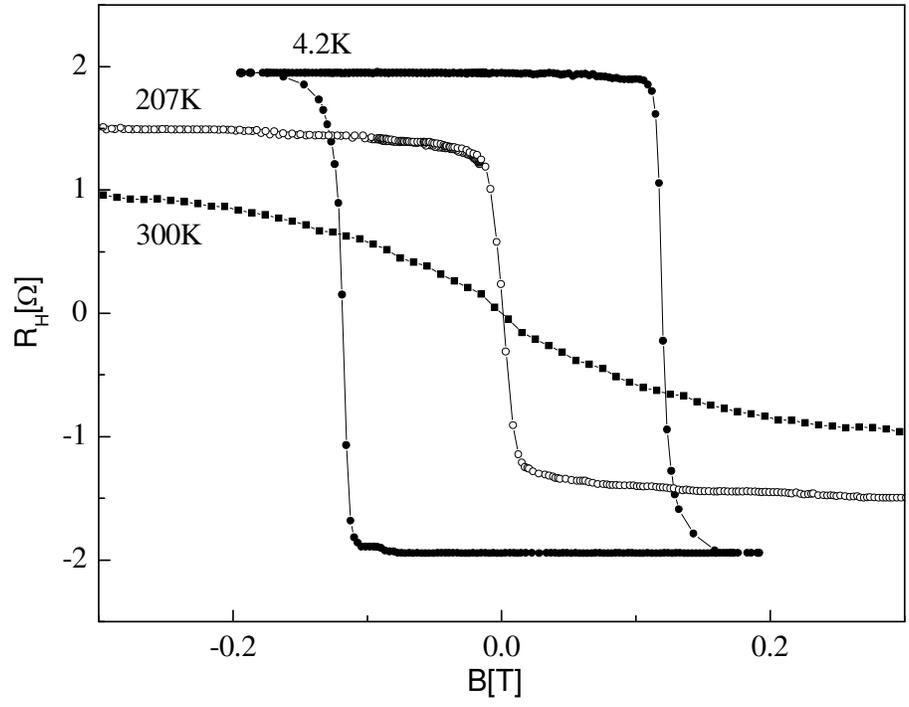

Fig. 6

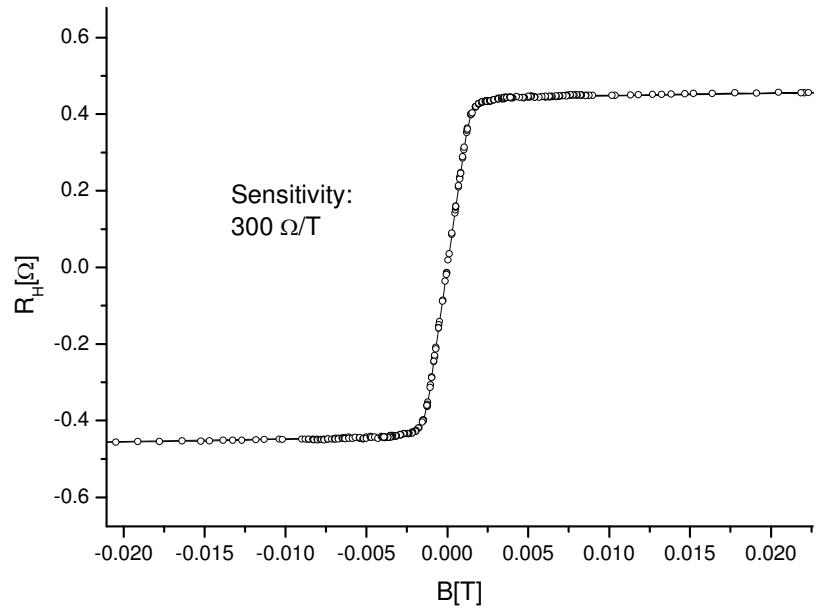

Fig.7.

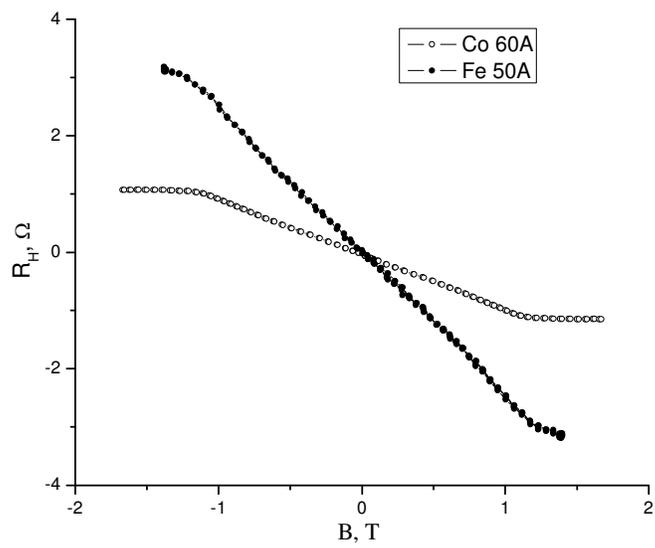

Fig. 8.

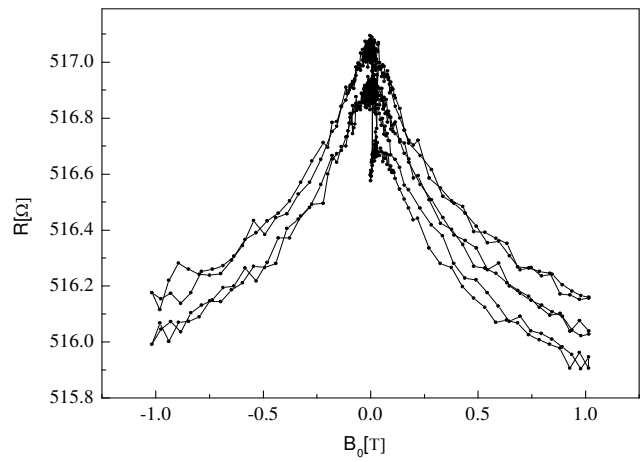

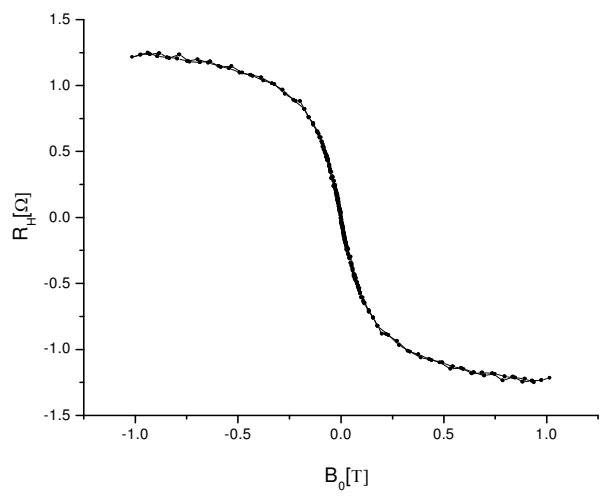

Fig. 9

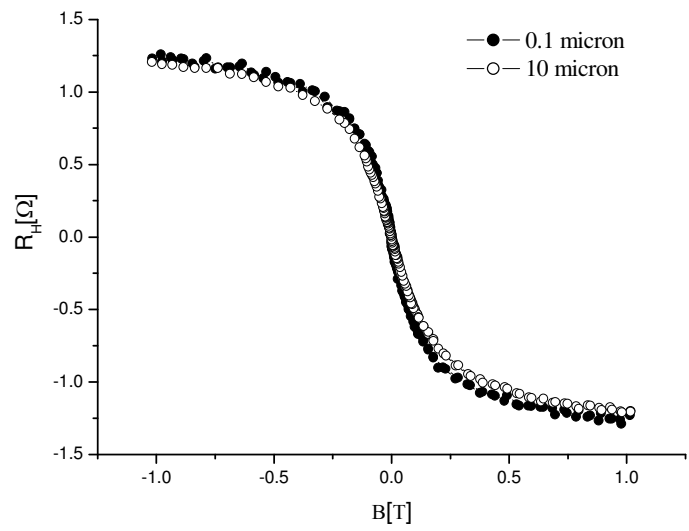

Fig.10

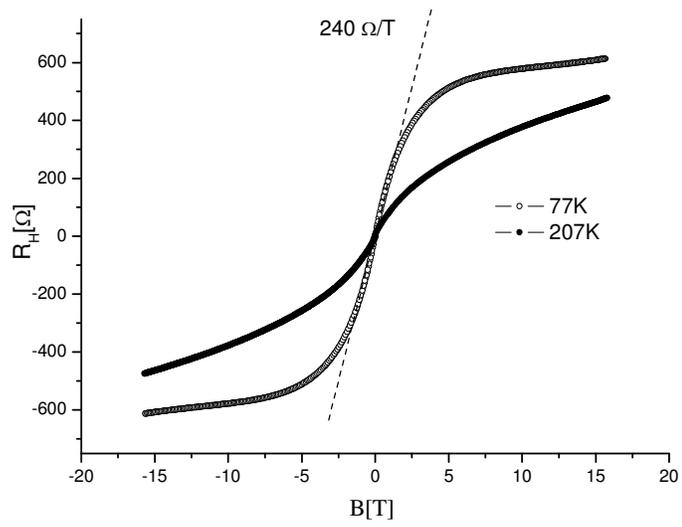

Fig. 11

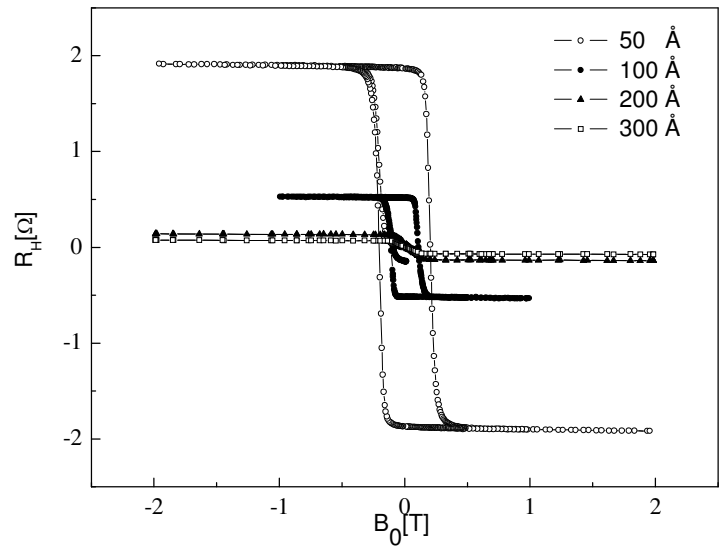

Fig. 12

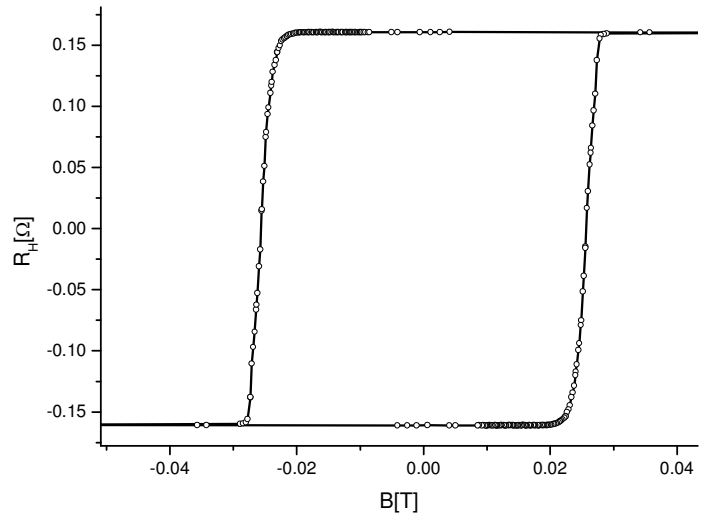

Fig. 13

.